\documentstyle[12pt]{article}

\begin{document}

 %%%%%%%%%%%%%%%%%%%%%%%%%%%%%%%%%%%%%%%%%%%%%%%%%%%%%%%%%%%%%%%%%%%
 % sugawara's macro
 %%%%%%%%%%%%%%%%%%%%%%%%%%%%%%%%%%%%%%%%%%%%%%%%%%%%%%%%%%%%%%%%%%%
 \newcommand{\Om}{\Omega}
 \newcommand{\df}{\stackrel{\rm def}{=}}
 \newcommand{\co}{{\scriptstyle \circ}}
 \newcommand{\de}{\delta}
 \newcommand{\lb}{\lbrack}
 \newcommand{\rb}{\rbrack}
 \newcommand{\rn}[1]{\romannumeral #1}
 \newcommand{\msc}[1]{\mbox{\scriptsize #1}}
 \newcommand{\dsp}{\displaystyle}
 \newcommand{\scs}[1]{{\scriptstyle #1}}

 \newcommand{\ket}[1]{| #1 \rangle}
 \newcommand{\bra}[1]{| #1 \langle}
 \newcommand{\vac}{| \mbox{vac} \rangle }

 \newcommand{\e}{\mbox{{\bf e}}}
 \newcommand{\va}{\mbox{{\bf a}}}
 \newcommand{\bc}{\mbox{{\bf C}}}
 \newcommand{\br}{\mbox{{\bf R}}}
 \newcommand{\bz}{\mbox{{\bf Z}}}
 \newcommand{\bq}{\mbox{{\bf Q}}}
 \newcommand{\bn}{\mbox{{\bf N}}}
 \newcommand {\eqn}[1]{(\ref{#1})}

 \newcommand{\cp}{\mbox{{\bf P}}^1}
 \newcommand{\n}{\mbox{{\bf n}}}
 \newcommand{\sbz}{\msc{{\bf Z}}}
 \newcommand{\sn}{\msc{{\bf n}}}

 \newcommand{\be}{\begin{equation}}\newcommand{\ee}{\end{equation}}
 \newcommand{\bea}{\begin{eqnarray}} \newcommand{\eea}{\end{eqnarray}}
 \newcommand{\ba}[1]{\begin{array}{#1}} \newcommand{\ea}{\end{array}}

 \newcommand{\cleqn}{\setcounter{equation}{0}}
 \makeatletter
 \@addtoreset{equation}{section}
 \def\theequation{\thesection.\arabic{equation}}
 \makeatother

 \def\np{Nucl. Phys. {\bf B}}
 \def\pl{Phys. Lett. {\bf B}}
 \def\mpl{Mod. Phys. {\bf A}}
 \def\ijmp{Int. J. Mod. Phys. {\bf A}}
 \def\cmp{Comm. Math. Phys.}
 \def\prd{Phys. Rev. {\bf D}}

 \def\ds{dS_{d_-,d_+}}
 \def\ads{AdS_{d_-,d_+}}
 \def\min{{\cal M}_{d_- +1,d_+}}
 \def\ds{dS_{d_-,d_+}}
 %% OJO!!!
 \def\g{G}

 \def\va{\vec a}
 \def\vb{\vec b}
 \def\vu{\vec u}
 \def\vv{\vec v}
 \def\vt{\vec t}
 \def\vn{\vec n}
 \def\ve{\vec e}
 \def\vx{{\vec x}}
 \def\vxM{{\vec x}_{+}}
 \def\vxm{{\vec x}_{-}}
 \def\vwM{{\vec w}_{+}}
 \def\vwm{{\vec w}_{-}}
 \def\vnM{{\check n}_{+}}
 \def\vnm{{\check n}_{-}}
 \def\dM{{d_{+}}}
 \def\dm{{d_{-}}}
 \def\ro{r_{0}{}}
 \def\vS{\vec {S}}
 \def\vsuno{{{\vec s}_1}}
 \def\vsdos{{{\vec s}_2}}
 % OJO!!!
 \def\ym{y}
 \def\hX{{\hat X}}
 \def\hJ{{\hat J}}
 \def\hP{{\hat P}}
 \def\hK{{\hat K}}
 \def\hD{{\hat D}}

 %%%%%%%%%%%%%% MACROS MONO %%%%%%%%%%%%%%

 \def\vr{{\check e}_r}
 \def\vteta{{\check e}_\theta}
 \def\vfi{{\check e}_\varphi}

 %%%%%%%%%%%%%%%%%%%%%%%%%%%%%%%%%%%%%%%%%
 \newcommand{\matriz}[4]{\left(
 \begin{array}{cc}#1 \\#3 \end{array}\right)}

 %%%%%%%%%%%%%%%%%%%%%%%%%%%%%%%%%%%%%%%%%%%%%%%%%%%%%%%%%%%%%%%%%%%
 \begin{flushright}
 La Plata Th-99/10\\September 7, 1999
 \end{flushright}

 \bigskip

 \begin{center}

 {\Large\bf Monopole and Dyon Solutions in AdS space}

 \bigskip
 \bigskip

 {\it \large Adri\'{a}n R.~Lugo\footnote{CONICET} \footnote
 {lugo@dartagnan.fisica.unlp.edu.ar}
 and Fidel A.~Schaposnik \footnote{Associate CICBA}}
 \bigskip

 {\it Departmento de F\'\i sica, Facultad de Ciencias Exactas \\
 Universidad Nacional de La Plata\\ C.C. 67, (1900) La Plata,
 Argentina}
 \bigskip
 \bigskip

 \end{center}

 \begin{abstract}
 We consider monopole and
 dyon classical solutions of the Yang-Mills-Higgs system coupled
 to gravity in
asymptotically
 anti-de Sitter space. We discuss both singular and regular solutions to
 the second order equations of motion showing that singular
 Wu-Yang like dyons can be
 found, the resulting metric being of the Reissner-N\"ordstrom type
 (with cosmological constant). Concerning regular solutions, we analyze the
 conditions under which they can be constructed discussing, for vanishing
 coupling constant, the main distinctive features related to the
 anti-de Sitter asymptotic condition; in particular, we find  in this case
 that the v.e.v. of the Higgs scalar, $|\vec H(\infty)|$,
 should be quantized in
 units of the natural mass scale $1/e\, r_0$ (related to the cosmological constant) according to
 $|\vec H(\infty)|^2 = m(m+1) (e\, r_0 )^{-2}$,
 with $m \in Z$.
\end{abstract}

\date{\today}

 %\newpage
\bigskip
\section{Introduction}
 \cleqn

Soon after the discovery of the magnetic monopole
and dyon   solutions \cite{tHP}-\cite{BPS} in spontaneously
broken $SU(2)$
gauge theory  with Higgs fields in the adjoint representation, different kinds
of self-gravitating monopoles were
discussed \cite{BR}-\cite{vNWP}.
Subsequently, the gravitational properties of these solutions
 and the relation between
monopoles and black holes were thoroughfully investigated
\cite{LNW}-\cite{BFM2}; more recenlty,  numerical studies clarifying gravitational instabilities and
other peculiar features of the solutions have been presented
\cite{BHK}-\cite{LW}
(see \cite{Gib}-\cite{M}
for a more complete list of references). All these investigations
correspond to asymptotically flat spaces.
Less is known when space time is
modified to include a cosmological constant $\Lambda$, in particular
for $\Lambda < 0$, i.e. the case in which space
is asymptotically anti-de Sitter space.
Recently, black hole solutions \cite{Wi}  and soliton solutions \cite{BH}
have been found in this case  for the  Einstein-Yang-Mills system
(no Higgs field).

It is the purpose of this work to study monopole and dyon
solutions for Yang-Mills-Higgs theory coupled to gravity
in asymptotically anti-de Sitter (AdS) space.

In Section 2 we
present the model,
the spherically symmetric ansatz for the gauge field, the Higgs triplet
and the metric, leading to radial equations of motion. Then,
in Section 3 we present a dyonic AdS black hole solution
(analogous to the one constructed in \cite{BR}-\cite{CF} in asymptotically
flat space).
Following different  perturbative approaches, we analyse in Section 4
regular monopole solutions and discuss its main properties. Finally, we
summarize our results in Section 5.

\section{The model, the Ansatz and the equations of motion}
\cleqn

We adopt the conventions in \cite{Wald}-\cite{Eguchi} for metric
curvature, gauge fields and else
\be
{\it F} \equiv d{\it A} + {\it A}\wedge {\it A}
\ee
where ${\it A}$ is a one-form connection.

A Higgs field $\; (H^i )\;$ transforms in the representation $R$ of the
semisimple gauge group $G$ which has Lie algebra {\it G} generated by $X_a$
with commutation relations and metric (that rises and lows indices)
\bea
[X_a , X_b] &=& f^c{}_{ab} \; X_c\cr
\kappa_{ab} &\equiv& <X_a ,X_b >
\eea
$D_\mu$ stands for general covariant derivative and $g_{ij}$ is a gauge
invariant metric.

The action we consider is
given by
\bea
S &=& S_G + S_{YM} + S_H = \int d^Dx\sqrt{|G|} ( L_G + L_{YM} + L_H )\cr
L_G &=& \frac{1}{\alpha_0}  \left( \frac{1}{2} R - \Lambda \right)\cr
L_{YM} &=& \frac{1}{4e{}^2}\; F_{\mu\nu}^a F_a^{\mu\nu} \cr
L_H &=& -\frac{1}{2} g_{ij}\; G^{\mu\nu}\; D_\mu H^i\; D_\nu H^j - V(H)
\label{1}
\eea
where $V(H)$ is the Higgs potential, $\alpha_0\equiv 8\pi G_D$ with
$G_D$ is the Newton constant in $D$ dimensions (to be taken
as $D = 4$ in what follows), $e$ the gauge coupling
and $\Lambda$ is the
cosmological constant (one easily sees that with our conventions
$\Lambda < 0$ corresponds, in the absence of matter, to anti-de Sitter
space).

The equations of motion that follow from (\ref{1}) are
\bea
E_{\mu\nu} + \Lambda\; G_{\mu\nu} &=&
\alpha_0 \; (T_{\mu\nu}^{YM} +  T_{\mu\nu}^H )\cr
g_{ij}\; {\tilde D}_\rho D^\rho H^j &=& \partial_i V(H)\cr
\frac{1}{e{}^2} D_\rho F_a^{\mu\rho} &=&
g_{ij}\; D^\mu H^i\; R(X_a)^j{}_k \; H^k
\eea
where the matter energy-momentum tensor is
\bea
T_{\mu\nu}^{YM} &=& \frac{1}{e{}^2}(- F_{\mu\rho}^a F_{a\;\nu}{}^\rho
+ \frac{1}{2}\; G_{\mu\nu}\; F_{\rho\sigma}^a F_a^{\rho\sigma} )\cr
T_{\mu\nu}^H &=& g_{ij} \; D_\mu H^i\; D_\nu H^j + G_{\mu\nu}\; L_H
\label{T}
\eea
and $\;{\tilde D}_\mu{}^i{}_j \equiv D_\mu \delta^i_j +\Gamma[g]^i_{jk}
D_\mu H^k\;$ contains the Christoffel symbol of $g_{ij}$.

Let us take now for definiteness $G= SU(2)$, $g_{ij} =\delta_{ij}$
for the Higgs fields in the
adjoint representation ``3", and a basis where $f^c{}_{ab}=\epsilon_{abc}$
with $\; < X_a , X_b >   \equiv \frac{1}{2} tr_3 X_a X_b = -\delta_{ab}$.
The most general static spherically symmetric form for the metric in $3$
spatial  dimensions togheter with the t'Hooft-Polyakov-Julia-Zee ansatz
for the gauge and Higgs fields in the usual vector notation reads
 \bea
 G &=& - \mu(x)\; A(x)^2\; d^2 t + \mu(x)^{-1}\; d^2 r + r^2\; d^2
\Omega_2\cr
\vec A &=& dt \; e\; h_0\; J(x)\; \vr - d\theta\; (1 - K(x) )\; \vfi
+ d\varphi \; (1 - K(x) )\; \sin\theta\; \vteta\cr
\vec H &=& h_0\; H(x)\; \vr
\label{ansatz}
\eea
 where we identify the constant $h_0$ (assumed non zero) with the minimum
 of the potential
 \be
 V(H) = \frac{\lambda}{4}\; ( {\vec H}^2 - h_0{}^2 )^2
 \ee
and we introduce the dimensionless coordinate $\; x\equiv e\; h_0\; r$.

Using this ansatz, the equations of motion take the form
\bea
\left( x\; \mu (x)\right)' &=& 1  + 3\; \gamma_0\;  x^2
- \alpha_0 h_0{}^2
\left( \mu (x)\; V_1 + V_2 + \frac{x^2}{2}\;\frac{J'(x)^2}{A(x)^2} +
\frac{J(x)^2 K(x)^2}{\mu(x) A(x)^2} \right)\cr
x\; A'(x) &=& \alpha_0 h_0{}^2 \left( V_1 + \frac{J(x)^2 K(x)^2}{\mu(x)^2
A(x)^2} \right) A(x) \label{eqq}
\eea
\bea
\left( \mu (x) A(x) K'(x) \right)' &=& A(x)\; K(x) \left( \frac{K(x)^2
-1}{x^2} + H(x)^2- \frac{J(x)^2}{\mu (x) A(x)^2} \right)\cr
\left( x^2\mu(x) A(x) H'(x) \right)' &=& A(x)\; H(x) \left( 2\; K(x)^2
+\frac{\lambda}{e{}^2}\; x^2\; ( H(x)^2 - 1) \right)\cr
\mu (x)\left( \frac{x^2 J'(x)}{A(x)} \right)' &=& \frac{2\; J(x)
K(x)^2}{A(x)}
\label{eqs}
\eea
where for convenience we have defined the dimensionless parameter
\be
\gamma_0 \equiv - \frac{\Lambda}{3 e{}^2 h_0{}^2}
\ee
  and
\bea
V_1 &=& K'(x)^2 + \frac{x^2}{2} H'(x)^2 \cr
V_2 &=& \frac{ (K(x)^2 -1)^2}{2\; x^2} +
\frac{\lambda}{4 e{}^2}\; x^2\; (H(x)^2 -1)^2
\eea

\subsection*{The boundary conditions}

Ansatz ({\ref{ansatz}) will lead to  well behaved solutions for the matter
fields if, at $x=0$, one imposes
\begin{itemize}

\item ${H(x)}/{x}$ and  ${J(x)}/{x}$ are regular;

\item $1- K(x)$ and $K'(x)$ go to zero.
\end{itemize}

On the other hand we want the system to go asymptotically to anti-de Sitter
space which corresponds to the solution of the Einstein equations with $\Lambda <0$
in absence of matter (see next Section); for
this to happen we must impose that the matter
energy-momentum tensor vanishes at spatial infinity.
>From eq.(\ref{T}) one can see that the appropriate conditions
 for $x\rightarrow\infty$
are
\bea
K(x) &\rightarrow& O(x^{-\alpha_1})\cr
H(x) &\rightarrow& H_\infty + O(x^{-1-\alpha_2})\cr
J(x) &\rightarrow& J_\infty + O(x^{-\alpha_3})
\label{boun}
\eea
with $\alpha_i > 0, ~i=1,2,3$.

\section{A dyonic AdS black hole in four dimensions}
\cleqn

Before discussing regular solutions to equations of motion (\ref{eqs})
satisfying the boundary conditions specified above, let us discuss
a singular solution which has very interesting properties. Indeed,
one can easily see that the restriction
\bea
K(x) &=& 0 \cr
H(x) &=& 1
\label{res}
\eea
leads to a singular solution   that exhibits a  Dirac
string starting at  $x=0$ and so the Abelian Dirac monopole character
of this solution,
 related to
a  Wu-Yang-like dyon  \cite{WY} which, for asymptotically flat space
was considered long ago \cite{BR}-\cite{CF}.
The equations for the metric are trivially solved by (\ref{res})
\bea
A(x) &=& 1\cr
\mu (x) &=&  1 + \gamma_0\; x^2 -  \frac{a}{x} +
\frac{\alpha_0 h_0{}^2 (1 + b^2 )}{2\; x^2}
\label{a}
\eea
leading  to a metric of the Reissner-Nordstr\"om
type (with cosmological constant). The constant $a$ in (\ref{a})
is related to the mass of this AdS charged black hole.

Concerning the electric potential
\be
J(x) = - \frac{b}{x} + J_\infty
\label{b}
\ee
With this, the electric ($Q_e$) and magnetic ($Q_m$) charges of the
solution are respectively
\begin{eqnarray}
Q_e \equiv -\frac{1}{4\pi} \; \int_{S^2} *F^r &=&  b \label{Qe}\\
Q_m \equiv -\frac{1}{4\pi} \; \int_{S^2}  F^r &=&  1
\label{Qm}
\end{eqnarray}
We then see that the parameter $b$ determines the electric charge while
$J_\infty$ sets the scale for $J$. In fact,
the metric function $\mu(x)$ (eq. (\ref{a})) can be rewritten in the form
\begin{equation}
\mu (x) =   1 + \gamma_0\; x^2 -  \frac{a}{x} +
\frac{\alpha_0 h_0{}^2 (Q_m^2 + Q_e^2 )}{2\; x^2}
\label{aa}
\end{equation}
making explicit the role of the (unit)
 magnetic and electric charges, the same as
in the $\Lambda = 0$
case \cite{BHK}. At the radii for which the metric function
vanishes, there will be
in general event horizons, as it happens in
asymptotically flat space. Now, the condition $\mu(x) = 0$ leads here to a quartic
algebraic equation for $x$,
\begin{equation}
\gamma_0 x^4 + x^2 - ax +  Z^2 = 0
\label{hor}
\end{equation}
where we have defined
\begin{equation}
Z^2 =  \sigma_0(Q_m^2 + Q_e^2)  \, , \;\;\;\; \; \;
\sigma_0 = \frac{\alpha_0 h_0{}^2}{2}
\label{Z}
\end{equation}
The explicit expression for the roots of this equation is not very illuminating.
One can see  that for $\gamma_0 \ll a^{-2}$ (or $|\Lambda| \ll (1/a^2)e^2h_0^2$)
the horizons have, qualitatively,
the same behavior as in the asymptotically flat case: there is a critical value
$Z^2_{crit}$  for $Z^2$ such that there are two horizons for
$Z^2 < Z^2_{crit}$ and two complex conjugate roots for $Z^2 > Z^2_{crit}$. The
value of $Z^2_{crit}$ and the two horizons
can be determined as a power expansion in $a^2\gamma_0$ \cite{R}.
Let us end this section by noting that, although the gauge field
is singular at the origin,
the solution can be considered regular as a black hole, so that the
singularity could be hidden behind a horizon.

\section{The system in AdS space}
\cleqn

\subsection*{Statement of the problem}

In the  $\;\alpha_0 h_0{}^2 \rightarrow 0\;$ limit the gravitational
equations decouple from the matter ones leading for the metric
 to the solution
\bea
A(x) &=& 1\cr
\mu(x) &=& 1 + \gamma_0 \; x^2 -\frac{a}{x}
\label{not}
\eea
which is nothing but the vacuum solution of the Einstein equations
with a cosmological constant (assumed negative), and corresponds
to a neutral Schwarzchild black hole in AdS space.
Concerning the integration constant $a$, it is
related to the mass of the black hole and
will be put to zero in what follows.
This metric, in turn, acts as a (AdS) background
with radius $r_0$,
\be
\; r_0 = \sqrt{-{3}/{\Lambda}}\;
\label{radius}
\ee
for the Yang-Mills-Higgs system.

For simplicity we start studying  eqs  (\ref{eqs})
in the BPS limit which corresponds to
${\lambda}/{e{}^2} = 0$
with $h_0$ fixed.
\bea
( \mu(x) K'(x) )' &=& K(x) \left( \frac{K(x)^2 -1}{ x^2} + H(x)^2
 - \frac{J(x)^2}{\mu (x)} \right)\cr
( x^2\mu(x) H'(x) )' &=& 2\; H(x)\; K(x)^2 \cr
\mu(x)\; ( x^2 J'(x) )' &=& 2\; J(x)\; K(x)^2
\label{bps}
 \eea

The total amount of matter $M$
can be associated with
 the generator of traslations in time, $\partial_t$
(which appears in the AdS algebra). It takes the form
(see for example \cite{Gib})
\be
M = \int_{\Sigma_t}  d^3 x\; \sqrt{ g^{(3)}}\; T_{00}
\ee
where $\; g^{(3)} \;$ is the determinant of
the induced metric on surfaces $\Sigma_t$ of
constant time $t$ with normal vector
$\; e_0 = \mu(x)^{-\frac{1}{2}} \partial_t \;$ and
$\; T_{00}\equiv e_0^{\,\mu}\; e_0^{\,\nu}\; T_{\mu\nu} = {T_{tt}}/{\mu (x)}$
is the local energy density as seen by an observer moving on the flux
lines of $\partial_t$.
For the spherically symmetric configuration we are considering, it takes the
form
\be
M = \frac{4\pi h_0}{e}\; \int_0^\infty\; dx\;
\frac{x^2}{( 1+\gamma_0\; x^2)^\frac{3}{2} }\;
\frac{T_{tt}}{e{}^2\;h_0{}^4}
\ee
We quote for completeness
the explicit expressions for $T_{tt} = T_{tt}^{(YM)} + T_{tt}^{(H)} $
\bea
\frac{T_{tt}^{(YM)}}{e{}^2\;h_0{}^4} &=&
\frac{\mu(x)}{2}\; J'(x)^2 + \frac{J(x)^2\;K(x)^2}{x^2} +
\frac{\mu(x)^2\; K'(x)^2}{x^2} + \frac{\mu(x)}{2\;x^4}\; ( K(x)^2-1)^2\cr
\frac{T_{tt}^{(H)}}{e{}^2\;h_0{}^4} &=&
\frac{\mu(x)^2}{2}\; H'(x)^2 + \frac{\mu(x)}{x^2}\; H(x)^2\; K(x)^2
\eea
It is not difficult to see from these expression
that the boundary conditions imposed through eqs.(\ref{boun})
are precisely those required for finiteness of $M$.

\subsection*{Attempts towards a perturbative solution}

In flat  Minkowski space  eqs.(\ref{bps})
become BPS  equations with the well-honoured Prasad-Sommerfield
solution saturating
the Bogomol'nyi bound \cite{BPS}
\bea
 K_0(x) &=& \frac{x}{\sinh x}\cr
 H_0(x) &=& \cosh\gamma\; f(x)\cr
 J_0(x) &=& \sinh\gamma\; f(x)
 \label{bpss}
 \eea
 where the constant $\gamma$ defines the boundary condition at infinity and
 \be
 f(x) \equiv \coth x - \frac{1}{x}
 \ee
For asymptotically flat space-times one can prove that self-gravitating
monopoles saturating the Bogomol'nyi bound do not exist
\cite{Gib}. In contrast, solutions to the second order equations of motion
can be found numerically,
 starting from the   flat space solution
(\ref{bpss}) and taking $\alpha_0$ sufficiently small. This regular
monopole (or dyon) solution ceases to exist for some critical value
$\alpha_0 = \alpha_0^c$ \cite{BFM1}-\cite{LW}.

One should expect that something similar happens in AdS space, at
least for small enough $\gamma_0$ (we are already in the $\alpha_0h_0^2 \to 0$ limit):
the existence of a small cosmological
constant  (with respect to the Higgs mass) should alter a
little the Prasad-Sommerfeld solution leading to a
dyonic
regular solution of the second order equations of motion in AdS space.
To analyse this possibility
one can try a perturbative solution around the Prasad-Sommerfield
solution in the form
 \bea
 K(x) &=& K_0(x) + \sum_{m\geq 1} \gamma_0{}^m K_m (x)\cr
 H(x) &=& H_0(x) + \sum_{m\geq 1} \gamma_0{}^m H_m (x)\cr
 J(x) &=& J_0(x) + \sum_{m\geq 1} \gamma_0{}^m J_m (x)
 \label{kbps}
 \eea
 Inserting these expansions in equations (\ref{bps}) and comparing
  $\gamma_0$ powers, one gets a recursive set of inhomogeneous linear
 second order differential equations which can be written
 as
 \bea
 \left( \begin{array}{c} \frac{K_m (x)}{x f(x)}\\
 \frac{H_m (x)}{f(x)} \\ \frac{J_m (x)}{f(x)} \end{array}\right) '' +
   2\; \left( \ln (xf(x))\right)'\left( \begin{array}{c} \frac{K_m (x)}{x
f(x)}\\
   \frac{H_m (x)}{f(x)} \\ \frac{J_m (x)}{f(x)}\end{array}\right) ' -
 V(x) \left( \begin{array}{c} \frac{K_m (x)}{x f(x)} \\
   \frac{H_m (x)}{f(x)} \\ \frac{J_m (x)}{f(x)}\end{array}\right) =
\frac{1}{x f(x)}\; {\vec q}_m (x)
%\left( \begin{array}{c} \frac{q_m^1 (x)}{x f(x)} \\
% \frac{q_m^2 (x)}{f(x)} \\ \frac{q_m^3 (x)}{f(x)}\end{array}\right)
\label{larg}
 \eea
for all $m\geq 1$. Here $V(x)$ is a $m$-independent matrix defined as
\bea
 V(x) &=& (2\; f(x)\;\coth x  -1 )\;
\left( \begin{array}{ccc} 1&0&0\\0&0&0\\ 0&0&0\end{array} \right)\cr
&+& \frac{2\; f(x)}{\sinh x}\;
\left( \begin{array}{ccc} 0&\cosh\gamma&-\sinh\gamma\\
2\cosh\gamma &0&0\\ 2\sinh\gamma&0&0\end{array} \right)
\eea
Concerning the
inhomogeneous term in the r.h.s. of (\ref{larg}),  vectors
${\vec q}_m$  can be determined at order $m$ from the
solution at lower orders from the closed expressions
\bea
q_m^1 &=& \sum_{p=0}^{m-1} (-x^2 )^{m-1-p}\; ( K_{p}(x) - 2\; x\;
K'_{p}(x))\cr
&+& \sum_{k,l,s=0}^{m-1} (-x^2)^{m-k-l-s}\; K_k (x) \big(
\frac{K_l (x) K_s (x) }{x^2} + H_l (x) H_s (x)\cr
&-& (m+1-k-l-s) J_l (x) J_s (x) \big) \cr
q_m^2 &=& 2\; \sum_{p=0}^{m-1} (-x^2 )^{m-p}\; H'_p (x) +
\frac{2}{x} \sum_{k,l,s=0}^{m-1} (-x^2)^{m-k-l-s}\; K_k (x) K_l (x) H_s
(x)\cr
q_m^3 &=& \frac{2}{x} \sum_{k,l,s=0}^{m-1} (-x^2)^{m-k-l-s}\;
K_k (x) K_l (x) J_s (x)
\label{otro}
\eea
(Sums above are restricted according to $\; k+l+s\leq m\;$).
One can easily finds  the explicit form in the $m=1$ case,
\be
{\vec q}_1 = \left( \begin{array}{c}
\frac{x^3}{\sinh x}\; \left( 1 - 2\; f(x)^2 + f(x)^2 \; \sinh^2\gamma
\right)\\
-2\cosh\gamma\; \left( 1 - 2\; \frac{x^2}{\sinh^2 x} +
\frac{x^3}{\sinh^3 x}\;\cosh x \right)\\
-2\; \sinh\gamma\; \frac{x^3\; f(x)}{\sinh^2 x}
\end{array}\right)
\ee
At each order $m$, eqs.(\ref{larg}) corresponds to a coupled linear
system of
three differential equations with three unknowns,
obeying boundary conditions defined by eqs.(\ref{boun}).
Now, this  system can be reduced  by diagonalizing  the
off diagonal part of $V(x)$;
indeed, defining $\; (h_m^1 , h_m^\pm )\;$  by
\bea
K_m (x) &=& \frac{1}{\sqrt 2}\; x\; f(x)\; h_m^+ (x) \cr
H_m (x) &=& f(x)\; ( \sinh\gamma\; h_m^1 (x) + \cosh\gamma\; h_m^- (x)
)\cr
 J_m (x) &=& f(x)\; ( \cosh\gamma\; h_m^1 (x) + \sinh\gamma\; h_m^- (x) )
 \eea
 for all $m\geq 1$, eqs.(\ref{larg}) become
 \bea
\left( x^2\; f(x)^2\; h_m^1{}'(x)\right)' &=& x\; f(x)\; (-\sinh\gamma\;
q_m^2 (x) + \cosh\gamma\; q_m^3 (x) ) \label{al}\\
 h_m^+{}''(x) + 2 \;\left(\ln(xf(x))\right)'\; h_m^+{}'(x) &=&
(2\; f(x)\coth x  -1 ) h_m^+ (x) +
 2\; {\sqrt 2}\; \frac{f(x)}{\sinh x}\; h_m^- (x)\cr
 &+& \frac{\sqrt 2}{xf(x)}\; q_m^1 (x) \label{bal}\\
 h_m^-{}''(x) + 2\; \left(\ln(xf(x))\right)'\; h_m^-{}'(x) &=&
 2\;{\sqrt 2}\;\frac{f(x)}{\sinh x}\; h_m^+ (x) \cr
&+& \frac{1}{xf(x)} \left( \cosh\gamma\; q_m^2 (x) -\sinh\gamma\; q_m^3
(x)\right)
\label{cal}
\eea
Identifying
$\;H_\infty \equiv\cosh\gamma\;$ and  $\;J_\infty \equiv \sinh\gamma\;$,
boundary conditions can be simply stated in terms of these functions:
solutions must be regular at
$x=0$ and vanish for  $x \to \infty$.

Eq. (\ref{al}) can be easily integrated, this
giving $h_m^1$; one is then left with the reduced system
(\ref{bal})-(\ref{cal}),
with two unknowns $\; h_m^\pm$.
Although we were not able to solve this system in closed form,
the  first order $m=1$ solution of eq.(\ref{al}), $\; h_1^1\; $,
 can be easily integrated
with the result
\be
h_1^1(x) = \frac{x}{f(x)}
\ee
which is well behaved in $x=0$ but grows linearly with $x$ at infinity!
This could be taken as a signal that a solution of the type proposed
in (\ref{kbps}) does
not exist,   indicating that the theory with
cosmological constant  strictly zero has a completely different
behavior from that with $\Lambda \ne 0$,
regardless how small $\Lambda$  could be.
Another indication in the same direction
comes from the comparison of the AdS boundary conditions (\ref{boun})
with those corresponding to the flat space case,
\bea
K(x) &\rightarrow& O(x^{-\alpha_1})\cr
H(x)&\rightarrow& H_\infty + O(x^{-\alpha_2}) \;\;\;\;\;\;\; \;\;\;\;\;\;\; x \gg 1 \cr
J(x) &\rightarrow& J_\infty + O(x^{-\alpha_3})
\eea
Comparing these flat space conditions with those
corresponding to AdS space (conditions (\ref{boun}))
we see that the latter require the Higgs field
to reach its vacuum expectation value  faster than in flat space.
It is seemingly that the unbounded behaviour of $\mu(x) - 1$
for $\gamma_0 \ne 0$ (eq.(\ref{not})) is at the
root of this situation.
It is worth to note at this point that away from the BPS limit
($\;{\lambda}/{g_0{}^2} \neq 0\;$) the vanishing of the Higgs potential
at infinity requires exactly the same boundary both in AdS and in flat space.

\subsection*{Re-statement of the problem and asymptotic expansions}

A hint about the possible
non regular behaviour of the
solution in terms of the parameter $\gamma_0$
comes from the following analysis.

If and only if $\gamma_0$ is different from zero and positive
(e.g. in the case of AdS
space)
one can introduce the following
functions
\bea
k(y) &\equiv& \sqrt{\mu(x)}\; K(x)|_{x=\frac{y}{\sqrt{\gamma_0}}}\cr
h(y) &\equiv& x\; \sqrt{\mu(x)}\; H(x)|_{x=\frac{y}{\sqrt{\gamma_0}}}\cr
j(y) &\equiv& x\; J(x)|_{x=\frac{y}{\sqrt{\gamma_0}}}
\label{change}
\eea
in term of which the equations of motion take the form
\bea
y^2\;(1+y^2 )^2\; k''(y) &=& k(y)\; \left( k(y)^2 + h(y)^2 - j(y)^2
-1\right)\cr
y^2\;(1+y^2 )^2\; h''(y) &=& 2\; h(y)\; \left( k(y)^2 + y^2\;(y^2 +
\frac{3}{2})\right)\cr
y^2\;(1+y^2 )^2 \; j''(y) &=&  2\; j(y)\;k(y)^2
\label{eq}
\eea
The dependence on $\gamma_0$ has completely disappeared
from (\ref{eq}) through the non
analytical change (\ref{change})

Note  the monopole mass  can be written as
\be
M =\frac{4\; \pi h_0}{e} f(0,\gamma_0) = \frac{4\; \pi}{e^2\; r_0}\; f_0
\ee
w%
here, extending the usual flat space
notation \cite{tHP} we have introduced the dimensionless function
$f = f(\lambda/e^2,\gamma_0)$ and which in the Prasad-Sommerfield
limit can be written as  $f(0,\gamma_0) = \sqrt{\gamma_0 }f_0$ with
$f_0$ a numerical constant which can be in principle calculated
and
is of course finite if the appropriate  boundary conditions  hold.
Thus, having  AdS space a natural scale $r_0$, the system
trades
(in the Prasad-Sommerfield limit)  $h_0$
for the AdS radius $r_0 = \sqrt{-3/\Lambda}$ which now
sets the scale for the mass.

Returning to eqs.(\ref{eq}), one should  note  that in the
 $\; y<< 1\;$ region, which in the original variable corresponds to
$\; x<< {1}/{\sqrt{\gamma_0}}\;$, one  could hope to find
the BPS
solution since $\gamma_0$ is effectively very small and the
domain is approximately flat; in fact it is easy to check that the system
reduces in this region to the BPS equations.
Therefore, the problem can be settled as follows: a finite mass monopole (or
dyon) in AdS space   should interpolate between the
BPS solution near $y=0$,
\bea
k(y) &=& K_0 (y)\cr
h(y) &=& y\; H_0 (y)\cr
j(y) &=& y\; J_0(y)
\eea
and a solution that asymptotically  behaves as
\bea
k(y) &\rightarrow& O(y^{1-\alpha_1})\cr
h(y) &\rightarrow& H_\infty \;y^2 + O(y^{1-\alpha_2})\cr
j(y) &\rightarrow& J_\infty\; y + O(y^{1-\alpha_3})
\eea
In the intermediate ($y \sim 1$) region  (\ref{eq})
corresponds to a constant coefficient system, this ensuring
the existence of a solution in the neighborhood, according to
standard theorems on non-linear differential
equations systems.

\subsection*{The pure magnetic monopole.}

Let us still simplify the problem by
considering neutral  ($j=0$) magnetic monopoles.
We try
at large distances
a power series solution of the form
\bea
k (y) &=& \sum_{n=1}^{\infty}\; k_n\; y^{-n}\, ,   \;\;\;\;\;\;  y \gg 1 \cr
h (y) &=&  H_\infty\; y^2 + \sum_{n=0}^{\infty}\; h_n\; y^{-n}
\, ,   \;\;\;\;\;\; y \gg 1\label{plug}
\eea
Plugging these expansions into  eqs.(\ref{eq}) one
gets a compatible set of
recursion relations.
It is worth to note that  this procedure does not work
in flat space: in order
system to close, it should neccessary to add also positive powers,
because of the exponential decay behaviour of the  solution (\ref{bpss}).
This signals a very different qualitative behaviour of the solution in
AdS with respect to flat space,
its origin being of course
the quadratically divergent behaviour of $\mu(x)$
in AdS space,
which produces a power six (instead of two) in the r.h.s.
of (\ref{eq}).

Coming back to the expansions, one gets
for for $k_n$ the recursive relations
\bea
(2 - H_\infty{}^2 ) \ k_1&=& 0\cr
(6 - H_\infty{}^2 ) \ k_2&=& 0\cr
(12 - H_\infty{}^2 ) \ k_3&=& (2\; h_0\; H_\infty - 4 )\; k_1\cr
(20 - H_\infty{}^2 ) \ k_4&=& 2\; H_\infty \; h_1\; k_1 +
(2\; h_0\; H_\infty - 12 )\; k_2\cr
(30 - H_\infty{}^2 ) \ k_5&=& (2\; h_2 \; H_\infty + h_0{}^2  - 3 )\;
k_1\cr
&+& 2\; H_\infty\; h_1\; k_2 + (2\; h_0\; H_\infty - 24 )\; k_3\cr
~ & & ~ \cr
\ldots  \;\;\;  \ldots \;\;\; \ldots
& \ldots & \ldots  \;\;\;  \ldots \;\;\;\ldots \;\;\; \ldots  \;\;\;
\ldots \;\;\;\ldots\;\;\; \ldots\cr
~ & &  ~\cr
\left( n(n+1) - H_\infty{}^2 \right)\; k_n &=& f_1^{(n)} \; k_1 + \;\;\; \;\;\;
\ldots
\;\;\;  \;\;\;+
f_{n-2}^{(n)} \; k_{n-2}
\label{puntos}
\eea
where the coefficients $f_i^{(n)}$ are determined by the coefficients
$k_k$ with  $k<n-2$ and also by coefficients $h_n$ (Although inspection of (\ref{puntos}) shows no dependence
on $k_k$ for  $n=1,\ldots, 5$,  the situation changes for
$n>5$).

It is convenient at this point
to  redefine $ h_1 = H_\infty\; A$. Then, the solution for
$h(y)$ takes the form
\be
h(y) = H_\infty\; \left( y^2 + \frac{1}{2} + \frac{A}{y} - \frac{1}{8\; y^2}
- \frac{A}{10\; y^3}
+ (\frac{1}{48} + \frac{k_1{}^2}{9})\; \frac{1}{y^4} - \frac{11}{280}\;
\frac{A}{y^5} + O(\frac{1}{y^6}) \right)
\label{free}
\ee
with $A$ a free parameter.
The analysis of the system (\ref{puntos}) shows up a remarkable feature.
One easily verifies that, unless
\be
H_\infty{}^2 = m \; (m + 1) \;\;\;\; , \;\;\; m = 1 , 2 , .....
\label{quan}
\ee
one has  $\; k(y)\equiv  0\;$.

Now,  when $\; k(y) =  0\;$ the gauge field corresponds to
a Dirac monopole configuration while  $h(y)$ is just
the solution of the linear equation
\be
h''(y) = \frac{3 + 2\; y^2}{(1+ y^2)^2}\; h(y)
\ee
given by
\be
h = y \sqrt{1+y^2}
\label{nol}
\ee
which corresponds to $H(y) = 1$. This shows that this solution coincides
with that discussed in Section 3.
(There is a second solution for  $h$ which has to be discarded
since it is not regular at the origin).

More interesting are the solutions for which $H_\infty$
is quantized
according to
(\ref{quan}), which implies that the Higgs field takes at infinity the
value (see eq.(\ref{ansatz}))
\be
\left \vert \vec H(\infty) \right \vert^2_\Lambda = m(m+1)\; \left( \frac{1}{e\, r_0} \right)^2
\label{ang}
\ee
to be compared with the flat space answer
\be
\left \vert \vec H(\infty) \right \vert^2_{\Lambda=0} =  h_0{}^2
\label{ang0}
\ee
Now, in the Prasad-Sommerfield flat space case, being the Higgs
potential absent, $h_0$ is introduced
both in order to set a scale and as a vestige of the vanishing
potential. This is achieved precisely through  the asymptotic condition
(\ref{ang0}) which is not imposed by the potential (as in the
$\lambda \ne 0$ case) but nevertheless results to be a consistent asymptotic
condition for monopoles without electric charge.
With a similar purpose we introduced $h_0$ in curved space
as a scale, finding  that the consistent
asymptotic value of the Higgs scalar is constrained by the
AdS geometry to obey (\ref{ang}) instead of (\ref{ang0}).
Moreover the natural scale for the Higgs v.e.v. is set by the cosmological constant.
The fact that the asymptotic conditions  change in AdS space is already
encountered for the simple case of a free massive field where one finds
that the scalar cannot approach a constant at infinity \cite {Wit}.
In the present case, where the scalar is coupled to a gauge field,
the Higgs scalar has to behave at infinity according to  (\ref{ang}) and
even the lowest possible non-trivial value for $m$, $m=1$,
gives in anti-de Sitter space a squared Higgs field v.e.v. which is
twice the flat space value measure in the natural units of each problem.
In some sense this behavior resembles, for the purely magnetic solution in AdS
space, to the consistent asymptotic condition for flat space Prasad-Sommerfield
dyons which is not  $h_0$ but $  h_0 \cosh \gamma$ \cite{BPS}.

A first analysis of eqs.(\ref{puntos})  indicates that
a non trivial monopole solution should exist in asymptotically anti-de
Sitter space. Indeed, $k(y)$ can be written
at large distances,
in the form
\be
k(y) = \frac{k_m}{y^m}\; \left( 1 - \frac{a_m}{y^2} + O({1}/{y^3})
\right)
\ee
where $a_m$ and  the higher order terms can
be  straighforwardly computed from (\ref{puntos}):
 $\; a_1 = 1/5 , a_2 = 3/7 , a_3 = 2/3,\dots\;$.
We see that $k(y)$ exhibits an  $y^{-m}$ behavior. The
corresponding coefficient
 as well as $A$ are
free parameters which should be in principle determined by
matching this behavior with
the solution for small $y$.

Concerning the case in which an electric field is
included, direct
inspection of the third equation in (\ref{eq})
reveals that an ansatz of the form
\be
j(y) = J_\infty \; y - b  + \frac{j_5}{y^5} + ...
\ee
together with the ansatz (\ref{plug}) for $ k(y)$ and $h(y)$ should work as well.
(The high $1/y$ power is dictated by the necessity of cancelling the $\mu(x)$ factor)

\section{Discussion}
We have discussed in this work monopole and dyon configurations
for Yang-Mills-Higgs theory coupled to gravity when a cosmological constant
is included so that space-time is, asymptotically, anti-de Sitter space. Making
the usual spherically symmetric ansatz to separate the equations
of motion, we have investigated both singular and
regular configurations carrying magnetic and electric charge.

Concerning singular solutions, we have constructed a Wu-Yang like dyon solution
with a metric of the Reissner-Nordstr\"om type (with cosmological constant).
The
event horizons can be determined as a power expansion in $a^2\gamma_0$, with
$a$ related to the mass of the black hole and $\gamma_0$ proportional to
the cosmological constant.
Although the gauge field is singular at the origin, the solution
can be considered regular as a black hole, with the singularity hidden
behind the horizon.

In order to find regular solutions to the coupled equations of motion, we have
investigated different regimes. With vanishing gravitational constant,
Einstein equations decouple from matter and the solution for the metric
corresponds to a neutral Schwarzchild black hole in AdS. This metric acts as a
background for the Yang-Mills-Higgs system, this changing radically
the properties of the solution with respect to the asymptotically flat space
case.

First, we have tried to find solutions to the second order equations of motion,
close to the BPS ones, which have been shown to exist in the small Newton
constant regime for asymptotically flat space.  Now,
when a cosmological constant is included
(no matter how small this constant is) no solution close to the BPS
configuration can be found. As we showed, it is the change
in the asymptotic behavior of the Higgs field due
to a non zero cosmological constant that prevents such a solution.

We have then studied the problem of matching the conditions that the solutions
have to satisfy in asymptotically AdS space with those required at the origin,
in order to have finite mass. Working for
simplicity in the Prasad-Sommerfield limit,  we have found a remarkable result:
the v.e.v. $|\vec H(\infty)|$ of the Higgs scalar should obey
$|\vec H(\infty)|^2 = m(m+1) (e\, r_0 )^{-2}$  with $m=1,2, \ldots$ (note that this result is
obtained when an asymptotic power series behaviour is assumed).
When this condition is fulfilled, our analysis shows that
a monopole solution can be constructed with a finite mass whose scale is
set by the AdS radius.
We leave the numerical analysis of such monopole and
dyon solution for a forthcoming work.

 \section*{Acknowledments}
This work is partially supported  by CONICET, CICBA and  ANPCYT (PICT 97/2285)
(Argentina). A.R.L. would like to thank ISAS (Trieste) where part of
this work was done as well as helpful discussions with L.~Bonora and
A.~Bressan.

 \end{document}